%% file: Radcor2000DES.tex
\documentclass[12pt]{article}
\input epsf.tex
\def\DESepsf(#1 width #2){\epsfxsize=#2 \epsfbox{#1}}
\newcommand\pubnumber{}
\newcommand\pubdate{\today}
\newcommand\hepnumber{hep-ph/0102031}

\def\csumb{Institute of Theoretical Science\\
University of Oregon, Eugene, OR 97403 USA}
\def\support{\footnote{Work supported by the
U.~S.~Department of Energy}} 

\def\Title#1{\begin{center} {\Large\bf #1 } \end{center}}
\def\Author#1{\begin{center}{ \sc #1} \end{center}}
\def\Address#1{\begin{center}{ \it #1} \end{center}}

\newcommand\pubblock{\rightline{\begin{tabular}{l} \pubnumber\\
     \pubdate\\ \hepnumber \end{tabular}}}
\newenvironment{Abstract}{\begin{quotation}  }{\end{quotation}}
\newenvironment{Presented}{\begin{quotation} \begin{center} 
             Presented at the\end{center}
      \begin{center}\begin{large}}{\end{large}\end{center} \end{quotation}}

\makeatletter
\def\section{\@startsection{section}{0}{\z@}{5.5ex plus .5ex minus
 1.5ex}{2.3ex plus .2ex}{\large\bf}}
\def\subsection{\@startsection{subsection}{1}{\z@}{3.5ex plus .5ex minus
 1.5ex}{1.3ex plus .2ex}{\normalsize\bf}}
\def\subsubsection{\@startsection{subsubsection}{2}{\z@}{-3.5ex plus
-1ex minus  -.2ex}{2.3ex plus .2ex}{\normalsize\sl}}

\renewcommand{\@makecaption}[2]{%
   \vskip 10pt
   \setbox\@tempboxa\hbox{\small #1: #2}
   \ifdim \wd\@tempboxa >\hsize     
       \small #1: #2\par          
     \else                        
       \hbox to\hsize{\hfil\box\@tempboxa\hfil}
   \fi}

 \def\citenum#1{{\def\@cite##1##2{##1}\cite{#1}}}
 
\newcount\@tempcntc
\def\@citex[#1]#2{\if@filesw\immediate\write\@auxout{\string\citation{#2}}\fi
  \@tempcnta\z@\@tempcntb\m@ne\def\@citea{}\@cite{\@for\@citeb:=#2\do
    {\@ifundefined
       {b@\@citeb}{\@citeo\@tempcntb\m@ne\@citea\def\@citea{,}{\bf ?}\@warning
       {Citation `\@citeb' on page \thepage \space undefined}}%
    {\setbox\z@\hbox{\global\@tempcntc0\csname b@\@citeb\endcsname\relax}%
     \ifnum\@tempcntc=\z@ \@citeo\@tempcntb\m@ne
       \@citea\def\@citea{,}\hbox{\csname b@\@citeb\endcsname}%
     \else
      \advance\@tempcntb\@ne
      \ifnum\@tempcntb=\@tempcntc
      \else\advance\@tempcntb\m@ne\@citeo
      \@tempcnta\@tempcntc\@tempcntb\@tempcntc\fi\fi}}\@citeo}{#1}}
\def\@citeo{\ifnum\@tempcnta>\@tempcntb\else\@citea\def\@citea{,}%
  \ifnum\@tempcnta=\@tempcntb\the\@tempcnta\else
  {\advance\@tempcnta\@ne\ifnum\@tempcnta=\@tempcntb \else\def\@citea{--}\fi
    \advance\@tempcnta\m@ne\the\@tempcnta\@citea\the\@tempcntb}\fi\fi}
\makeatother

\input econfmacros2.tex

\begin{document}
\begin{titlepage}
\pubblock

\vfill
\def\thefootnote{\fnsymbol{footnote}}
\Title{Letting Real-Virtual Cancellations Happen by \\[5pt] 
Themselves in QCD Calculations}
\vfill
\Author{Davison E.~ Soper\support}
\Address{\csumb}
\vfill
\begin{Abstract}
Calculations of observables in quantum chromodynamics are typically
performed using a method that combines numerical integrations
over the momenta of final state particles with analytical integrations
over the momenta of virtual particles. I review a method for performing
all of the integrations numerically. In this method, the real-virtual
cancellations happen inside the integrals --  simply because they are
built into the Feynman rules. I indicate promising topics for further
research on this subject.
\end{Abstract}
\vfill
\begin{Presented}
5th International Symposium on Radiative Corrections \\ 
(RADCOR--2000) \\[4pt]
Carmel CA, USA, 11--15 September, 2000
\end{Presented}
\vfill
\end{titlepage}
\def\thefootnote{\arabic{footnote}}
\setcounter{footnote}{0}

\section{Introduction}

There is an important class of computer programs that do calculations in
quantum chromodynamics (QCD) in which the calculation is performed at
next-to-leading order in perturbation theory and allows for the
determination of a variety of characteristics of the final state.
This talk reviews a program of this class in which a ``completely
numerical'' integration algorithm is used.

I consider the calculation of ``three-jet-like'' observables in $e^+
e^-$ annihilation. A program that does this can be used
to calculate a jet cross section (with any infrared safe choice of jet
definition)  or observables like the thrust distribution. Such a program
generates random partonic events consisting of three or four final state
quarks, antiquarks, and gluons. Each event comes with a calculated 
weight. A separate routine then calculates the contribution to the
desired observable for each event, averaging over the events with their
weights.

The weights are treated as probabilities. However, these weights can be
both positive or negative. This is an almost inevitable consequence of
quantum mechanics. The calculated observable is proportional to the
square of a quantum amplitude and is thus positive. However, as soon
as one divides the amplitude into pieces for purposes of calculation,
one finds that, while the square of each piece is positive, the
interference terms between different pieces can have either sign. Thus
the kind of program discussed here stands in contrast to the tree-level
event generators in which, by simplifying the physics, one can
generally arrange to have all the weights be positive, or, even, all
be equal to 1.

To understand the algorithms used in the class of programs described
above, it is best to think of the calculations as performing
integrations over momenta in which the quantum matrix elements and the
measurement functions form the integrand. There are two basic algorithms
for performing the integrations. The older is due to Ellis, Ross, and
Terrano (ERT) \cite{Ellis:1981wv}. In this method, some of the
integrations are performed analytically ahead of time. The other
integrations are performed numerically by the Monte Carlo method. The
integrations are divergent and are regulated by analytical continuation
to $3 - 2\epsilon$ space dimensions and a scheme of subtractions or
cutoffs. The second method is much newer
\cite{Soper:1998ye,Soper:2000xk,beowulfcode}. In this method, all of the
momentum integrations are done by Monte Carlo numerical integration. With
this method, the integrals are all convergent (after removal of the
ultraviolet divergences by a straightforward renormalization procedure). 

In its current incarnation, the numerical method is not as good as older
programs in analyzing three jet configurations that are close to being
two jet configurations. On the other hand,   the numerical method offers
evident advantages in flexibility to modify the integrand. Since
this method is quite new, one cannot yet say for what problems it might
do better than the now standard ERT method. 

The numerical integration method exists as computer code with
accompanying technical notes \cite{beowulfcode} and many of the basic
ideas behind it have been described in two papers
\cite{Soper:1998ye,Soper:2000xk}. In this talk, I briefly review the
basics of the numerical integration method. Then, I display some graphs
that illustrate the cancellation of singularities that occurs inside the
integrand in the numerical method. Finally, I discuss some avenues for
future research.

\section{Review of the numerical method}
\label{sec:review}

Let us begin with a precise statement of the problem. We consider an
infrared safe three-jet-like observable in $e^+e^- \to {\it hadrons}$,
such as a particular moment of the thrust distribution. The observable
can be expanded in powers of
$\alpha_s/\pi$,
\begin{equation}
\sigma = \sum_n 
\sigma^{[n]},
\hskip 1 cm
\sigma^{[n]} \propto \left(\alpha_s / \pi\right)^n\,.
\end{equation}
The order $\alpha_s^2$ contribution has the form
\begin{eqnarray}
\sigma^{[2]} &=&
{1 \over 2!}
\int d\vec p_1 d\vec p_2\
{d \sigma^{[2]}_2 \over d\vec p_1 d\vec p_2}\
{\cal S}_2(\vec p_1,\vec p_2)
\nonumber\\
&&+
{1 \over 3!}
\int d\vec p_1 d\vec p_2 d\vec p_3\
{d \sigma^{[2]}_3 \over d\vec p_1 d\vec p_2 d\vec p_3}\
{\cal S}_3(\vec p_1,\vec p_2,\vec p_3)
\label{start}\\
&&
+
{1 \over 4!}
\int d\vec p_1 d\vec p_2 d\vec p_3 d\vec p_4\
{d \sigma^{[2]}_4 \over d\vec p_1 d\vec p_2 d\vec p_3 d\vec p_4}\
{\cal S}_4(\vec p_1,\vec p_2,\vec p_3,\vec p_4).
\nonumber
\end{eqnarray}
Here the $d\sigma^{[2]}_n$ are the  order $\alpha_s^2$ contributions
to the parton level cross section, calculated with zero quark masses.
Each contains momentum and energy conserving delta functions. The $d
\sigma^{[2]}_n$ include ultraviolet renormalization in the
$\overline{\rm MS}$ scheme. The functions $\cal S$ describe the
measurable quantity to be calculated. We wish to calculate a
``three-jet-like'' quantity.  That is, ${\cal S}_2 = 0$. The
normalization is such that ${\cal S}_n = 1$ for $n = 2,3,4$ would give
the order $\alpha_s^2$ perturbative contribution the the total cross
section.  There are, of course, infrared divergences associated with
Eq.~(\ref{start}). For now, we may simply suppose that an infrared
cutoff has been supplied.

The measurement, as specified by the functions ${\cal S}_n$, is to be
infrared safe, as described in Ref.~\cite{Kunszt:1992tn}: the ${\cal
S}_n$ are smooth, symmetric functions of the parton momenta and
\begin{equation}
{\cal S}_{n+1}(\vec p_1,\dots,\lambda \vec p_n,(1-\lambda)\vec p_n)
= 
{\cal S}_{n}(\vec p_1,\dots, \vec p_n)
\end{equation}
for $0\le \lambda <1$. That is, collinear splittings and soft
particles do not affect the measurement.

It is convenient to calculate a quantity that is dimensionless. Let the
functions ${\cal S}_n$ be dimensionless and eliminate the remaining
dimensionality in the problem by dividing by $\sigma_0$, the total
$e^+ e^-$ cross section at the Born level. Let us also remove the
factor of $(\alpha_s / \pi)^2$. Thus, we calculate
\begin{equation}
{\cal I} = {\sigma^{[2]} \over \sigma_0\ (\alpha_s/\pi)^2}.
\label{calIdef}
\end{equation}

Let us now see how to  set up the calculation of ${\cal I}$ in a
convenient form. We note that ${\cal I}$ is a function of the c.m.\ energy
$\sqrt s$ and the $\overline{\rm MS}$ renormalization scale $\mu$. We will
choose $\mu$ to be proportional to $\sqrt s$: $\mu = A_{UV} \sqrt s$.
Then ${\cal I}$ depends on $A_{UV}$. But, because it is dimensionless, it
is independent of $\sqrt s$. This allows us to write
\begin{equation}
{\cal I} = \int_0^\infty d \sqrt s\ h(\sqrt s)\ 
{\cal I}(A_{UV},\sqrt s),
\label{calI}
\end{equation}
where $h$ is any function with
\begin{equation}
\int_0^\infty d \sqrt s\ h(\sqrt s) = 1.
\label{rtsintegral}
\end{equation}

\begin{figure}
\centerline{\DESepsf(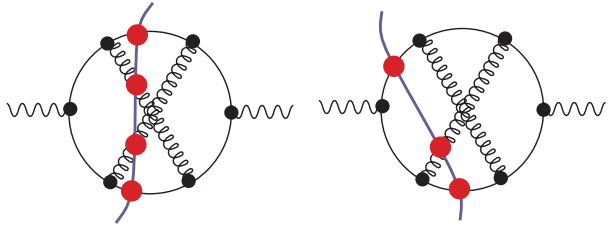 width 8 cm)}
\medskip
\caption{Two cuts of one of the Feynman diagrams that contribute to 
$e^+e^- \to {\it hadrons}$.}
\label{fig:cutdiagrams}
\end{figure}

The quantity $\cal I$ can be expressed in terms of cut Feynman
diagrams, as in Fig.~\ref{fig:cutdiagrams}. The dots where the parton
lines cross the cut represent the function ${\cal S}_n(\vec p_1,
\dots ,\vec p_n)$. Each diagram is a three loop diagram, so we have
integrations over loop momenta $ l_1^\mu$, $ l_2^\mu$ and
$ l_3^\mu$. We first perform the energy integrations. For the graphs
in which four parton lines cross the cut, there are four mass-shell
delta functions $\delta (p_J^2)$. These delta functions eliminate the
three energy integrals over $ l_1^0$, $ l_2^0$, and $ l_3^0$ as
well as the integral (\ref{calI}) over $\sqrt s$. For the
graphs in which three parton lines cross the cut, we can eliminate the
integration over $\sqrt s$ and two of the $ l_J^0$ integrals. One
integral over the energy $E$ in the virtual loop remains. We perform
this integration by closing the integration contour in the lower half
$E$ plane. This gives a sum of terms obtained from the original
integrand by some simple algebraic substitutions. Having performed the
energy integrations, we are left with an integral of the form
\begin{equation}
{\cal I} = 
\sum_G
\int d\vec  l_1\,d\vec  l_2\,d\vec  l_3\
\sum_C\,
g(G,C;\vec l_1,\vec l_2,\vec l_3).
\label{master}
\end{equation}
Here there is a sum over graphs $G$ (of which one is shown in
Fig.~\ref{fig:cutdiagrams}) and there is a sum over the possible cuts
$C$ of a given graph. The problem of calculating ${\cal I}$ is now set
up in a convenient form for calculation.

If we were using the Ellis-Ross-Terrano method, we would put the sum
over cuts outside of the integrals in Eq.~(\ref{master}).
For those cuts $C$ that  have three partons in the final state, there
is a virtual loop. We can arrange that one of the loop momenta, say
$\vec l_1$, goes around this virtual loop. The essence of the ERT
method is to perform the integration over the virtual loop momentum
analytically ahead of time. The integration is often ultraviolet
divergent, but the ultraviolet divergence is easily removed by a
renormalization subtraction. The integration is also typically infrared
divergent. This divergence is regulated by working in $3 - 2\epsilon$
space dimensions and then taking $\epsilon \to 0$ while dropping the
$1/\epsilon^n$ contributions (after proving that they cancel against
other contributions). After the $\vec l_1$ integration has been
performed analytically, the integrations over $\vec  l_2$ and $\vec
 l_3$ can be performed numerically. For the cuts $C$ that have four
partons in the final state, there are also infrared divergences. One
uses either a `phase space slicing' or a `subtraction' procedure to get
rid of these divergences, cancelling the $1/\epsilon^n$ pieces against
the $1/\epsilon^n$ pieces from the virtual graphs. In the end, we are
left with an integral $\int d\vec  l_1\,d\vec  l_2\,d\vec  l_3$ in
exactly three space dimensions that can be performed numerically.

In the numerical method, we keep the sum over cuts $C$ inside the
integrations. We take care of the ultraviolet divergences by simple
renormalization subtractions on the integrand. We make certain
deformations on the integration contours so as to keep away from poles
of the form $1/[E_F - E_I + i\epsilon]$. Then the integrals are all
convergent and we calculate them by Monte Carlo numerical integration.

Let us now look at the contour deformation in a little more detail. We
denote the momenta $\{\vec l_1,\vec l_2,\vec l_3\}$ collectively
by $ l$ whenever we do not need a more detailed description. Thus
\begin{equation}
{\cal I} = 
\sum_G
\int\! d l\
\sum_C\,
g(G,C; l).
\label{basicagain}
\end{equation}
For cuts $C$ that leave a virtual loop integration, there are
singularities in the integrand of the form $E_F - E_I +i\epsilon$ (or
$E_F - E_I -i\epsilon$ if the loop is in the complex conjugate
amplitude to the right of the cut). Here $E_F$ is the energy of the
final state defined by the cut $C$ and $E_I$ is the energy of a
possible intermediate state. These singularities do not create
divergences. The Feynman rules provide us with the $i\epsilon$
prescriptions that tell us what to do about the singularities: we should
deform the integration contour into the complex $l$ space so as to
keep away from them. Thus we write our integral in the form
\begin{equation}
{\cal I} = 
\sum_G
\int\! d l\
\sum_C\,
{\cal J}(G,C; l)\,
g(G,C; l+i\kappa(G,C; l)).
\label{deformed}
\end{equation}
Here $i\kappa$ is a purely imaginary nine-dimensional vector that we add
to the real nine-dimensional vector $ l$ to make a complex
nine-dimensional vector. The imaginary part $\kappa$ depends on the
real part $ l$, so that when we integrate over $ l$, the
complex vector $ l + i\kappa$ lies on a surface, the integration
contour, that is moved away from the real subspace. When we thus deform
the contour, we supply a jacobian ${\cal J} = \det(\partial ( l +
i\kappa)/\partial  l)$. (See Ref.~\cite{Soper:2000xk} for details.) 

The amount of deformation $\kappa$ depends on the graph $G$ and,
more significantly, the cut $C$. For cuts $C$ that leave no virtual
loop, each of the momenta $\vec  l_1$, $\vec  l_2$, and $\vec
 l_3$ flows through the final state. For practical reasons, we want
the final state momenta to be real. Thus we set $\kappa = 0$ for cuts
$C$ that leave no virtual loop. On the other hand, when the cut $C$
does leave a virtual loop, we choose a non-zero $\kappa$. We must,
however, be careful. When $\kappa = 0$ there are singularities in $g$
on certain surfaces that correspond to collinear parton momenta. These
singularities cancel between $g$ for one cut $C$ and $g$ for another.
This cancellation would be destroyed if, for $ l$ approaching the
collinear singularity, $\kappa = 0$ for one of these cuts but not for
the other. For this reason, we insist that for all cuts $C$, $\kappa \to
0$ as $ l$ approaches one of the collinear singularities. The details
can be found in Ref.~\cite{Soper:2000xk}. 

Much has been left out in this brief overview, but we should now have
enough background to see how the method works.

\section{Example}

\begin{figure}[htb]
\centerline{\DESepsf(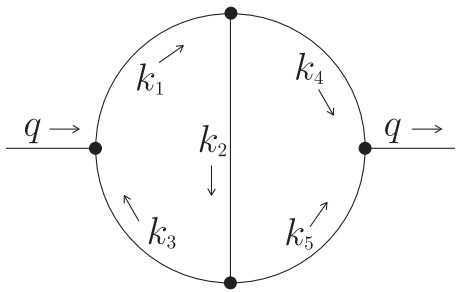 width 6 cm)}
\caption{Sample graph in $\phi^3$ theory.}
\label{fig:samplegraph}
\end{figure}

I present here a simple example, taken from Ref.~\cite{Soper:2000xk}.
Instead of working with QCD at three loops with many graphs, let's work
with one graph for $\phi^3$ theory at two loops, as shown in
Fig.~\ref{fig:samplegraph}.

\begin{figure}[htb]
\centerline{\DESepsf(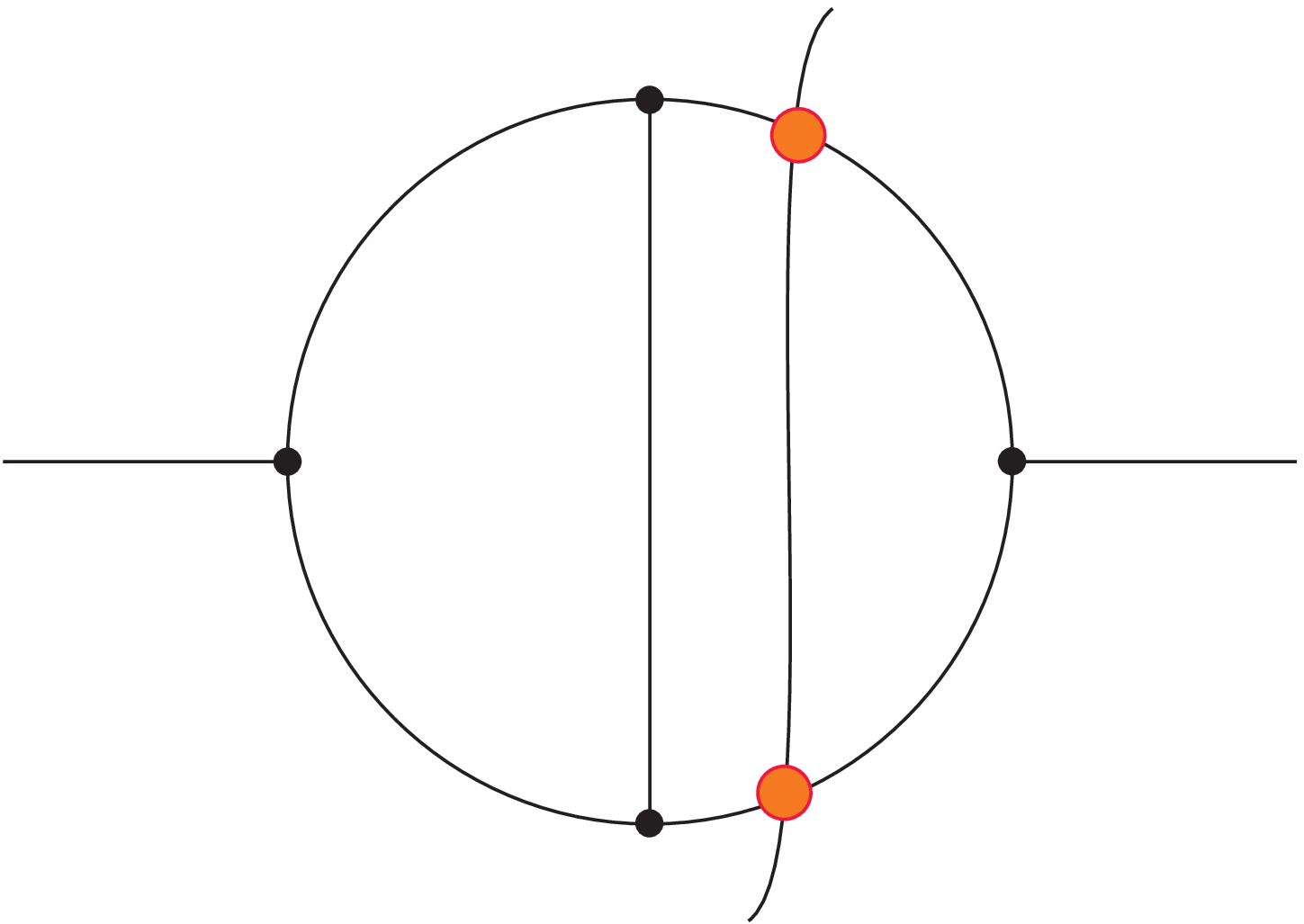 width 3 cm)
            \DESepsf(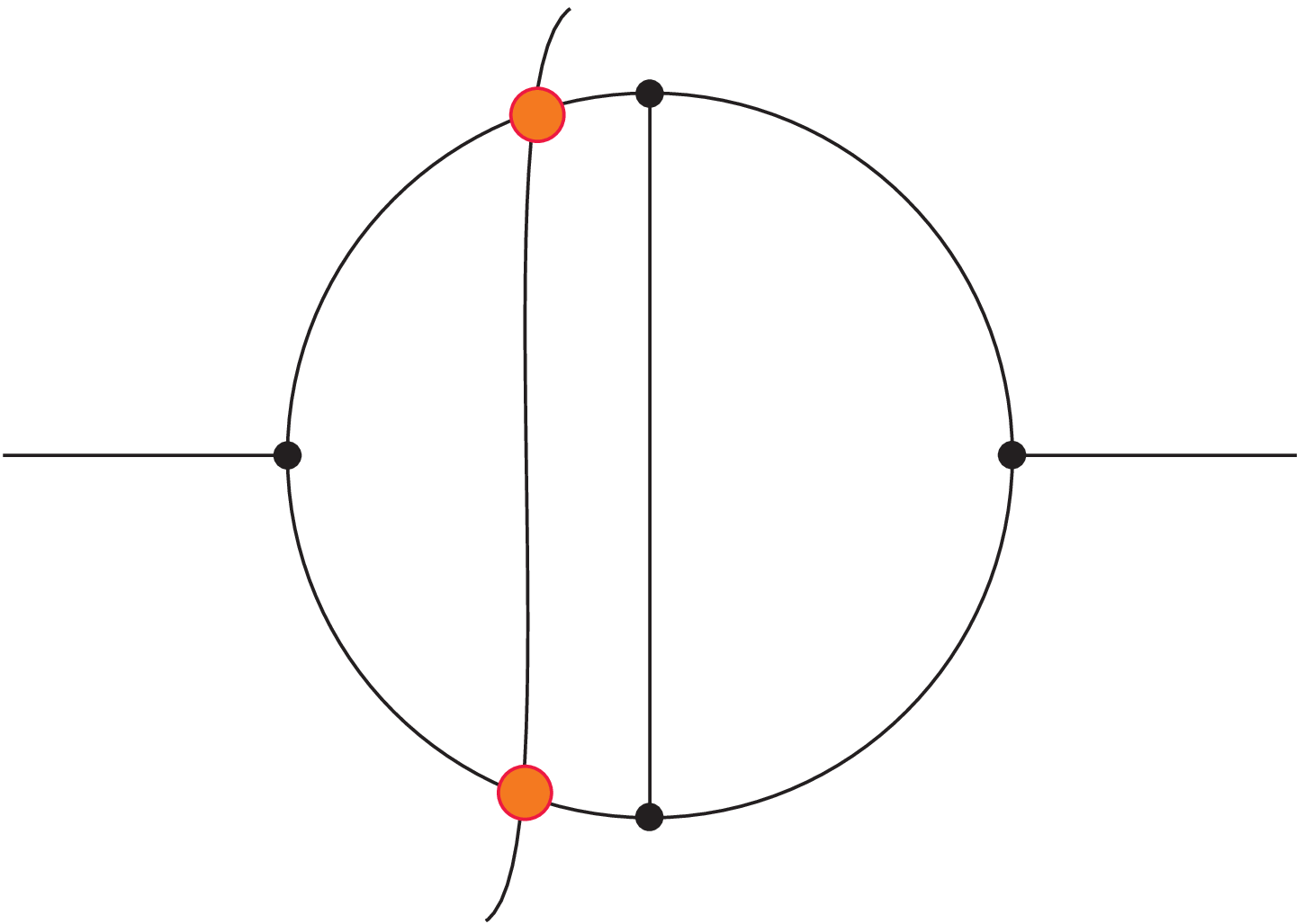 width 3 cm)}

\centerline{\DESepsf(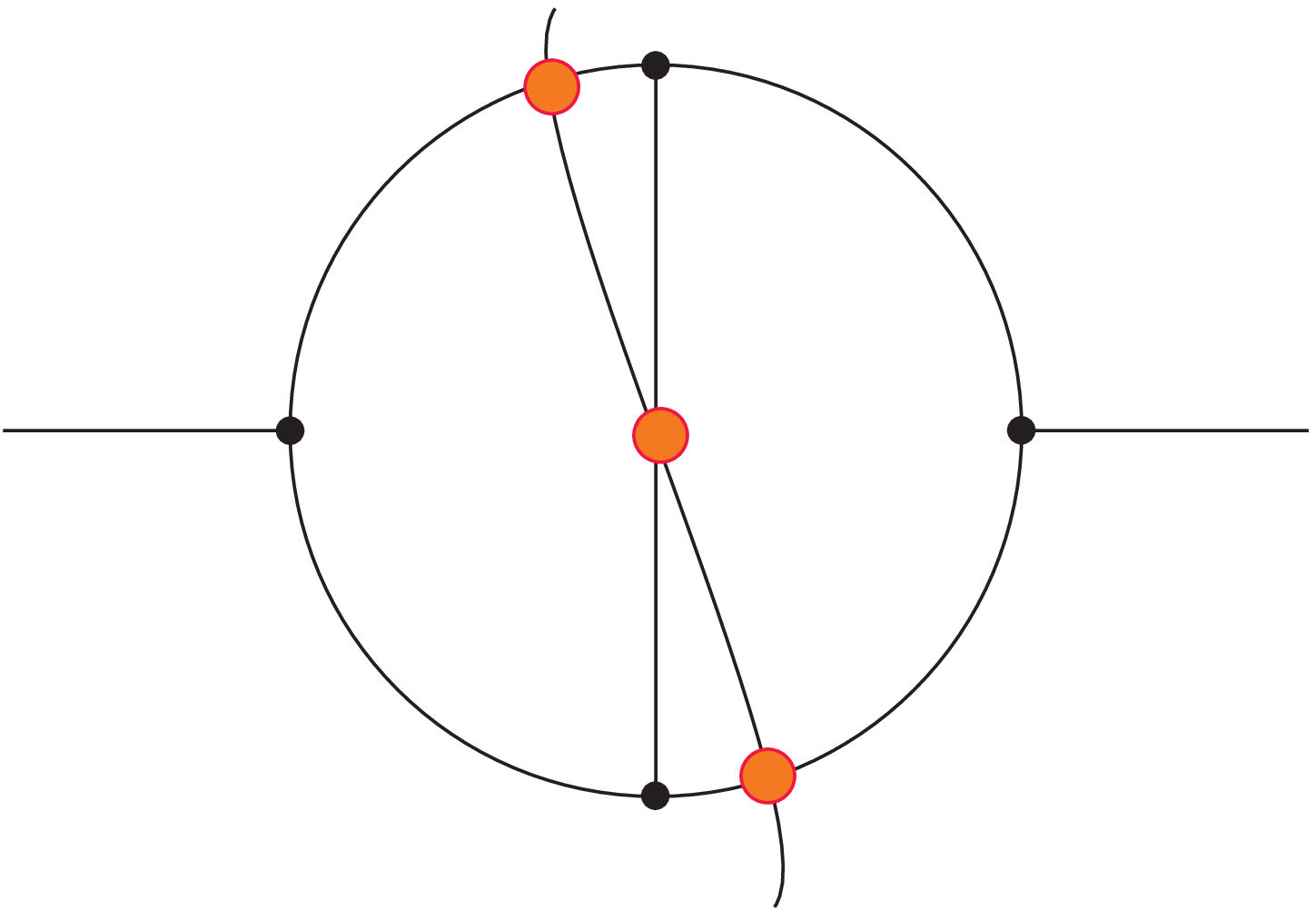 width 3 cm)
            \DESepsf(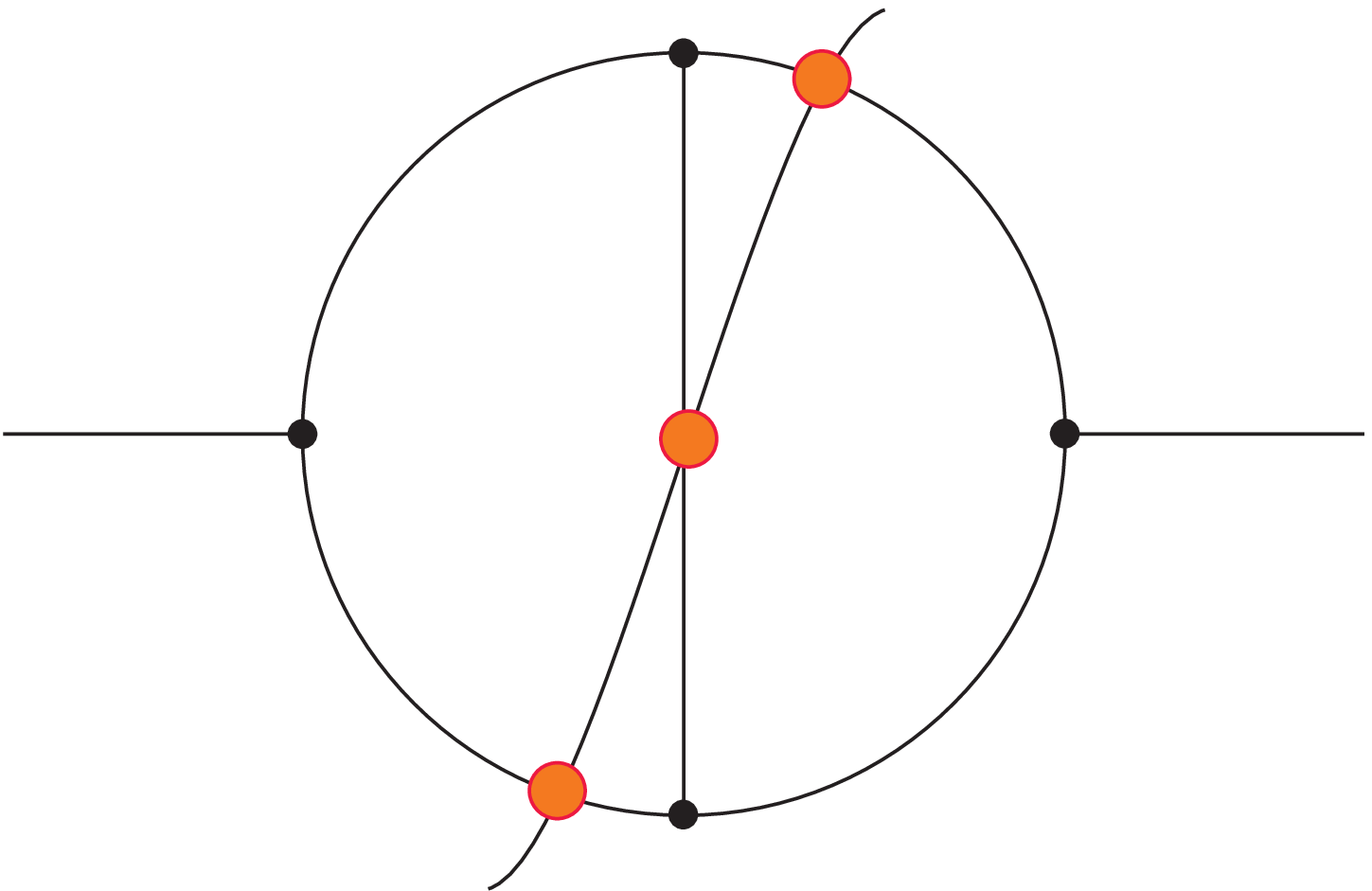 width 3 cm)}

\caption{Cuts of the sample graph.}
\label{fig:samplecuts}
\end{figure}
This graph has four final state cuts, as shown in
Fig.~\ref{fig:samplecuts}. We will fix the incoming momentum $\vec q$ and
integrate over the incoming energy $q^0$. For a measurement function, we
take ${\cal S}(p) = \sum |\vec p_{T,i}|$, where $\vec p_T$ is the part of
$\vec p$ orthogonal to $\vec q$. We make a choice of contour deformations
and of the density $\rho$ of Monte Carlo integration points as described
in \cite{Soper:2000xk}. Then we can plot the integrand $f$ divided by the
density of points $\rho$ versus the loop momentum. In a Monte Carlo
integration, large $f/\rho$ corresponds to large fluctuations, so
$f/\rho$ should never be too large. In the two figures that follow, I
plot $f/\rho$ versus the momentum in the left hand loop. Specifically,
using the $\vec k_n$ defined in Fig.~\ref{fig:samplegraph}, I plot$f/\rho$
versus $\vec l \equiv \vec k_2$ at fixed $\vec k_4$ for $\vec l$ in the
$\{\vec k_4,\vec q\}$ plane.

\begin{figure}[htb]
\centerline{\DESepsf(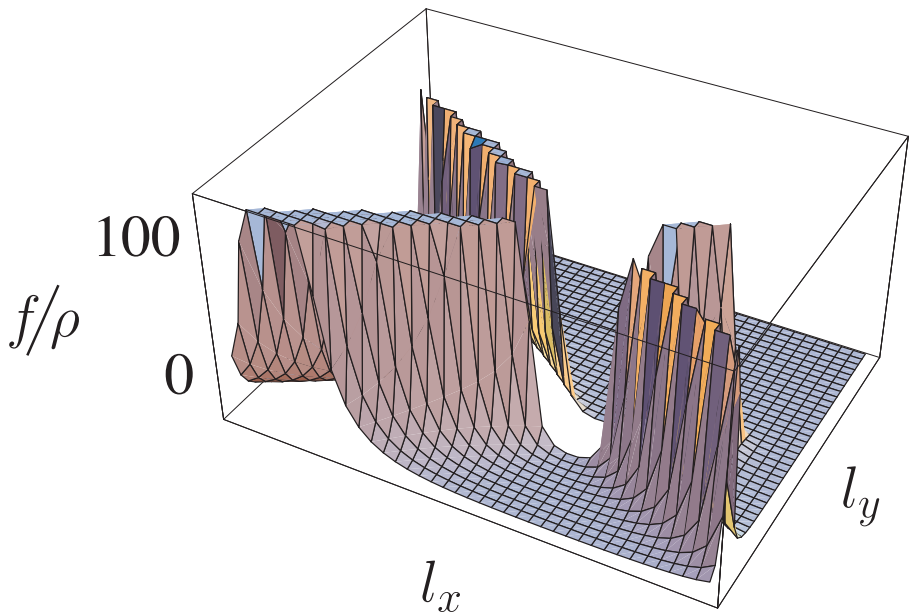 width 12 cm)}
\caption{Integrand divided by the density of points for the three parton
cuts. The collinear singularities are visible.}
\label{fig:real}
\end{figure}

\begin{figure}[htb]
\centerline{\DESepsf(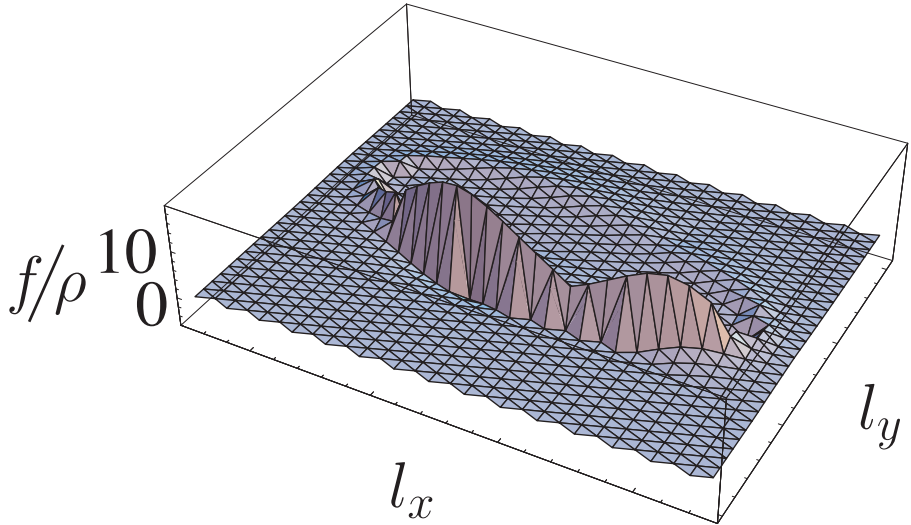 width 12 cm)}
\caption{Integrand divided by the density of points for all cuts
together. The collinear singularities disappear while the soft parton
singularity in $f$ is weakened so that it can be cancelled by a
singularity in $\rho$.}
\label{fig:total}
\end{figure}

In Fig.~\ref{fig:real}, I show $f/\rho$ summed over the two cut Feynman
graphs that have three partons in the final state, leaving no
virtual loop. Evidently, there are singularities. There is a soft parton
singularity (at $\vec l = 0$) that I have cut out of the diagram and
there are collinear parton singularities that are visible in the picture.
In the Ellis-Ross-Terrano method, these cut graphs would be calculated
using a numerical integration. But first a cutoff or other method for
eliminating the singularities would be needed to eliminate the singular
region. The two cuts that leave virtual subgraphs also lead to
singularities along the collinear lines in the space of the loop
momentum. I omit displaying a graph of $f/\rho$ for these two cut Feynman
graphs because the result simply looks like an upside down version of
Fig.~\ref{fig:real}. In the Ellis-Ross-Terrano method, one takes care of
the singularities in the virtual loop by integrating in $3 - 2 \epsilon$
space dimensions. In the numerical method, one combines the integrands
for all of the cuts. Then the collinear singularities disappear, while
the soft singularity is weakened enough that it can be eliminated in
$f/\rho$ by building a suitable singularity into $\rho$.  As suggested by
the title for this talk, the cancellation of singularities between real
and virtual graphs happens by itself because it is built into the Feynman
rules.  The result for $f/\rho$ summed over all four cuts is shown in
Fig.~\ref{fig:total}. The collinear singularities are gone, while the
soft parton singularity in $f$ has been weakened enough that it is
cancelled by a corresponding singularity in $\rho$. Thus a Monte Carlo
integration of $f$ using a density of integration points $\rho$ can
converge nicely because $f/\rho$ is not singular.

What remains visible in Fig.~\ref{fig:total} is a ridge in
$f/\rho$ for $\vec l$ lying on the ellipsoidal surface defined by $|\vec
k_1| + |\vec k_3| = |\vec k_4| + |\vec k_5|$, where the intermediate state
energy in the virtual graphs matches the final state energy. This ridge is
related to an energy denominator factor $1/[E_F - E_I + i\epsilon]$ in old
fashioned perturbation theory. The numerical integration method has taken
advantage of the $i\epsilon$ prescription in the Feynman rules to deform
the integration contour and avoid the singularity.

\section{Prospects}

There are a number of promising areas for further research along these
lines. 

The current program, {\it beowulf} \cite{beowulfcode}, does $e^+ + e^-
\to 3\ {\rm jets}$ at next-to-leading order. The partons are all massless.
With some modifications, the partons could have masses. Then one could
include massive quarks and one could extend the theory to the complete
Standard Model with its massive vector bosons. Furthermore, one could add
supersymmetry interactions.

The current program is confined to processes with no hadrons in the
initial state. Presumably the same idea can be applied to processes with
initial state hadrons, that is electron-proton collisions and
proton-proton or proton-antiproton collisions. Again, one should be able
to make the particles massive so that one can extend the calculations to
the complete Standard Model and supersymmetry.

It should also be possible to have more final state partons. That is, one
could attempt to calculate $e^+ + e^- \to 4\ {\rm jets}$ or $p + \bar p
\to  3\ {\rm jets}$ at next-to-leading order. 

The challenge of the legendary hero Beowulf was to kill the monster
Grendel. The monsters listed above are already dead or at least gravely
injured. In particular, all that {\it beowulf} can do could already have
been accomplished by the program of Kunszt and Nason eleven years ago
\cite{KN}. However, the challenge of calculating $e^+ + e^- \to 3\ {\rm
jets}$ at next-to-next-to-leading order remains unmet, and it may be that
a completely numerical attack would be successful.

A less difficult goal is to use the flexibility inherent in the numerical
method to go beyond fixed order perturbation theory. For instance, one
could use running couplings inside the next-to-leading order graphs as a
method for investigating power suppressed (``renormalon'') contributions
to the theory. More importantly, one could put a next-to-leading
order calculation inside a parton shower event generator (or attach
parton showers to the outside of the next-to-leading order calculation)
in order to have a full parton shower event generator that is correct at
next-to-leading order for three jet quantities in $e^+e^-$ annihilation.
This is, of course, not quite trivial \cite{DESMK}. As far as I can see,
the first step is to convert the current algorithms so that they operate
in Coulomb gauge instead of Feynman gauge. In this way, the partons
propagating into the final state have physical polarizations only. Then
these physically polarized partons can split many times to make parton
showers. One simply has to avoid counting the same splittings twice.


\end{document}

%% file: econfmacros2.tex
\def\beq{\begin{equation}}
\def\eeq#1{\label{#1}\end{equation}}
\def\eeqn{\end{equation}}

\newenvironment{Eqnarray}%
   {\arraycolsep 0.14em\begin{eqnarray}}{\end{eqnarray}}
\def\beqa{\begin{Eqnarray}}
\def\eeqa#1{\label{#1}\end{Eqnarray}}
\def\eeqan{\end{Eqnarray}}

\let\bar=\overbar

\def\Dslash{\not{\hbox{\kern-4pt $D$}}}
\def\dslash{\not{\hbox{\kern-2pt $\del$}}}

\def\msb{{\bar{\ssstyle M \kern -1pt S}}}

\def\lsim{\mathrel{\raise.3ex\hbox{$<$\kern-.75em\lower1ex\hbox{$\sim$}}}}
\def\gsim{\mathrel{\raise.3ex\hbox{$>$\kern-.75em\lower1ex\hbox{$\sim$}}}}